\documentstyle[11pt,aaspp4,psfig]{article}

\def\msol   {{M$_{\odot}$}}
\def\mo     {{M$_{\odot}$}}
\def\kms    {~km~s$^{-1}$}

\def\lax    {${_<\atop^{\sim}}$}
\def\aql    {Aql~X-1}

\slugcomment{Accepted for publication in ApJ}

\lefthead{Garcia et al.}
\righthead{Aql X-1 in Outburst and Quiescence}

\begin{document}

\title{Aql X-1 in Outburst and Quiescence}

\author{Michael R. Garcia\altaffilmark{1}, 
Paul J. Callanan\altaffilmark{2}}
\author{John McCarthy\altaffilmark{2},  
Kristoffer Eriksen\altaffilmark{1}, and Robert M. Hjellming\altaffilmark{3}}


\altaffiltext{1}{Harvard-Smithsonian Center for Astrophysics, MS-4, 60
Garden St., Cambridge, MA 02138; email: mgarcia,keriksen@cfa.harvard.edu}
\altaffiltext{2}{Physics Department, University College, Cork,
Ireland; email: paulc,mccarthy@ucc.ie}
\altaffiltext{3}{National Radio Astronomy Observatory, Socorro, NM
87801, U.S.A.; email: rhjellmi@aoc.nrao.edu}


\begin{abstract}

        We present photometry and spectroscopy of the soft x-ray
transient \aql .  Optical photometry during an active state shows a
strong (0.6~mag peak-to-peak) modulation at a period of 
19~hours.  Infrared (K$'$-band) photometry during a quiescent state
limits any ellipsoidal variations to $<0.07$~mag (peak-to-peak),
which implies an inclination $i < 31^{\rm o}$ (90\% limit).  
Spectroscopy in a quiescent
state shows at most very small radial velocity variations, which
implies a very low inclination of $i <12^{\rm o}$ (90\% limit).  
The low inclination is
rather unexpected given the large photometric modulation seen in the
active state.  The upper limit to the
equivalent width of the anomalous Li~6707\AA\/ line is $<0.3$\AA,
which is similar to the measured strength of this line 
in several other x-ray transients.

\end{abstract}




%

\section{Introduction}

Aql X-1 is the most active of the soft x-ray transients (SXTs), having
outbursts nearly once per year (Kaluzienski et~al. 1977, Priedhorsky
\& Terrell 1984). The accreting object is known to be a neutron star
(NS) due to the detection of type~1 x-ray bursts (Czerny, Czerny \&
Grindlay 1987).  During outbursts the optical counterpart (V1333 Aql)
brightens by several magnitudes to $V \sim 17$ or occasionally $V \sim
15$.  In-between outbursts the system fades to $V \sim 19$, and the
spectrum shows K0-3~V absorption features as well as emission lines
typical of low-mass x-ray binaries (LMXBs, Thorstensen, Charles \&
Bowyer 1978, Shahbaz, Casares \& Charles 1997). The reddening and
distance to the optical counterpart have variously been estimated at
${\rm E(B-V)} \sim 0.4$ or $\sim 0.35$, and $d \sim
2.5$~kpc or 2.3~kpc, based on the observed quiescent
colors and spectral type (Thorstensen, Charles \& Bowyer 1978,
Shahbaz et~al. 1996, Shahbaz et~al. 1998a).

A photometric modulation at $\sim 19$~hr is sometimes found during
 quiescence and mini-outburst. 
 SXT often show modulations at the orbital period during quiescence due
 to ellipsoidal effects, and in outburst due to x-ray heating of the
 accretion disk and/or the secondary (van Paradijs and McClintock 1995). 
Interestingly, there are two disparate determinations of the
 period: Shahbaz et~al. (1998b) find ${\rm P}=19.30 \pm 0.05$~hr, while
 Chevalier and Ilovaisky (1998) find ${\rm P}=18.9479 \pm 0.0002$~hr.
This latter modulation was first  found during outburst (Chevalier and
Ilovaisky 1991) and appears to be stable over several outburst cycles.
Recent quiescent data appear to show a modulation at the same period,
 and the phase of the quiescent and outburst minima are in agreement
(Chevalier and Ilovaisky 1998).  If this result is confirmed, 
the modulation is very likely at the orbital period.  The former modulation
was found over a 7~day period when the system was quiescent.  Effects which
could produce periodicities offset from the orbital period (ie, outburst
 superhumps) should therefore be absent, and the modulation should be
at the orbital period.  Therefore the discrepancy between the two periods is
hard to understand.

Shahbaz et~al. (1998b) measure a very small ellipsoidal modulation at
I-band during quiescence, and use this to estimate the inclination
$20^{\rm o} < i < 31^{\rm o}$.  However, they also caution that a nearby
comparison star shows a similar (but smaller) modulation at the same
period, and that their Aql X-1 measurement may also be interpreted as
an upper limit of $i < 31^{\rm o}$.  Shahbaz et~al. (1997) measure the
rotational velocity of the secondary star to be ${\rm V_{rot}~sin}(i) =
62^{+30}_{-20}$~\kms, and also note that the secondary contributes
94\% of the quiescent light at 6000\AA.

The x-ray decay lightcurve, and
changes in the x-ray spectrum during the 1997 February-March outburst
provide good evidence that the propeller effect dominates the
accretion flow in Aql X-1 at x-ray luminosities below $\sim 10^{36}$
erg~s$^{-1}$ (Campana et~al. 1998, Zhang, Yu and Zhang 1998).

The rapid outburst cycle makes Aql X-1 an ideal test object for
theories of SXT outbursts.  One possible explanation for the rapid
outburst cycle of Aql X-1 is offered by van Paradijs (1996), who
points out that Aql X-1 lies at the boundary of the (in)stability line
of the accretion disk limit cycle (Smak 1983).

The dominance of the secondary during quiescent intervals provides an
opportunity to determine the evolutionary state of the system, the
orbital period, and the mass function.  A determination of the mass of
the neutron star (which requires accurate knowledge of the mass
function and the inclination) is of particular interest, as the
current inventory of neutron star mass measurements is based largely
on massive systems (van Kerkwijk, van Paradijs, \& Zuiderwijk 1995).
Given that the neutron stars in low-mass systems have different
evolutionary histories, and may have formed via different methods
(accretion induced collapse is often mentioned, van Paradijs
et~al. 1997), there is a possibility their masses may differ.
Three low-mass systems in which NS masses have been measured are
PSR 1012+5307 (Callanan, Garnavich \& Koester 1998), Cen X-4 (Shahbaz,
Naylor and Charles 1993), and Cyg X-2
(Casares, Charles \& Kuulkers 1998).  While the NS in the former two 
systems have a masses consistent with the canonical 1.4\mo, the latter
may have a mass $>1.88$\mo.

\section{Observations}

\subsection{Outburst Photometry at the FLWO 1.2m:}

On the nights of 17, 18, and 19~July 1996 (UT), we obtained V-band
photometry of \aql\/ from the Fred Lawrence Whipple Observatory (FLWO)
1.2m telescope and `1-shooter' CCD camera.  The XTE/ASM rates show
that Aql X-1 was in a flat-topped, weak outburst for an $\sim 80$~day
period in June-July 1996.  The mean flux as measured by the ASM
was $\sim 18$~mCrab, but the rate is clearly variable.

The data were de-biased and flat-fielded with IRAF, and magnitudes
were determined with the aid of IRAF/DAOPHOT.  As observing conditions
were non-photometric, our magnitude zero point has been set via stars
h and k from Lyutyi \& Shugarov (1979)

This photometry folded on the 18.9479~hour period of Chevalier and
Ilovaisky (1998) is shown in Figure~1.  There is a marked modulation with a
peak-to-peak amplitude of $\sim 0.6$~magnitudes.  While the data do
not provide interesting constraints on the period, they do
determine the time of minimum light ${\rm T(0)} = 2450282.220 \pm 0.003$~HJD.

\subsection{Quiescent Photometry at the KPNO 2.1m:}

We used the IRIM on the KPNO 2.1m on the nights of 8-12 July 1997
(UT), to obtain J,H, and  K$'$-band photometry of V1333~Aql.  This imager
uses a HgCdTe, NICMOS3 device which provides 256x256 pixels of $1.1''$
size on the 2.1m.  Conditions were variable, but often photometric
with $\sim 1.5''$ seeing.  We observed the Aql X-1
field using a 3x3 dither pattern with variable $\sim 10''$ slews
between each dither.  At each position we exposed for 10~seconds
using 6~co-adds. The background was derived using the median of the 9
images and subtracted from each frame.  Each were then flat-fielded,
registered and co-added. Both the image reduction and photometry (see
below) were performed using IRAF/DAOPHOT.  We note that the
background images show systematic variations which appear to be
caused by the combination of the large pixel size and the relatively
crowded field.  These systematics are included in our errors.
The magnitude scale was set via observations of various stars from the
UKIRT Faint Standard catalogue (Casali and Hawarden 1993).

The K$'$-band lightcurve for Aql X-1 and a nearby (presumed constant)
star of similar magnitude is shown in Figure~2.  Each light curve has
been folded on the orbital period determined by Chevalier \& Ilovaisky
(1998) of ${\rm P}=18.9479$~hr, using a the T(0) determined above for phase
zero. The mean magnitudes for Aql X-1 were ${\rm J}=16.50\pm 0.15$,
${\rm H}=15.90\pm 0.15$, and K$'=15.74 \pm 0.15$.  These errors are
indicative of the observed scatter in the Aql X-1 K$'$ observations,
and include the variations found in repeated observations of the
standards over the run.

We then binned these two datasets into 12~equally sized phase bins,
determining the magnitude error for each bin from the observed
variance in the bin.  These data where then fitted to models of the
expected ellipsoidal lightcurve, using the model of Avni (1978) as
described by Orosz and Bailyn (1997).  This code computes the 
orbital variations
in the projected area of the star in the Roche geometry, 
along with limb and gravity darkening, in order to 
predict the lightcurve.  
For q (ratio of secondary mass
to primary mass) between 1 and 0.1, the best fit inclination ranges
between $10^{\rm o} < i < 12^{\rm o}$.  The 68\% error range includes
zero and is bounded by $i < 22^{\rm o}$; the 90\% error range is
bounded by $i<31^{\rm o}$.  Similar fits to the presumed constant star
show minima within a few degrees of zero, and 68\% and 90\% upper
limits at $i<22^{\rm o}$ and $i<30^{\rm o}$ respectively.  We note
that mean variability in the binned data is higher for the standard
star than that for Aql X-1: 7.3\% vs. 6.4\%.  This indicates that the
variability is dominated by systematic effects, likely due to the
large ($\sim 1''$) pixel size and the very crowded Aql X-1 field.

\subsection{MMT Spectroscopy:} 

We obtained 26 spectra of \aql\/ on the nights of 7,8, and 9 June
1997, using the MMT and Blue Channel spectrograph. 
The spectra cover the wavelength range 3600\AA~--~7200\AA\/ with 
3.6\AA\/ resolution.  Conditions were clear and seeing was $\sim 1''$.  
Exposure times were 20~minutes, and each exposure was bracketed
by an exposure of a HeNeAr comparison lamp in order to reduce the
spectra to a common velocity.  The dispersion is
1.2\AA~pixel$^{-1}$, and the S/N in each continuum pixel is $\sim 5$. 
Aql X-1 was near quiescence at this time, as evidenced by 
photometry with the WO 1.2m telescope which finds ${\rm V}=18.68 \pm 0.03$
on June 7, and non-detection
by the XTE ASM.

The images were reduced with standard IRAF routines, and spectra were
extracted with the IRAF longslit package.  Velocities were
determined via cross-correlations computed with
 the IRAF/FXCOR package run  against a range of late-type
standard stars.  We found that a late K star produced the strongest 
correlations, and therefore used GJ~9698 (spectral type
K8V, ${\rm HCV} = -28.2$~\kms, Bopp \& Meredith 1986) in subsequent velocity
determinations.  We limited our correlations to the wavelength ranges
4900\AA~--~5850\AA\/ and 5915\AA~--~6470\AA\/ in order to avoid the
possible ISM contribution in the NaD line and the H$\alpha$ and
H$\beta$ emission lines.  The more heavily reddened SXT GS~2000+25
shows weak interstellar lines at 5778\AA, 6177\AA, and 6284\AA\/
(Harlaftis, Horne, \& Filippenko 1996), but we find our results are
unchanged if we shorten our wavelength intervals to exclude these
lines.

In order to estimate the errors we also extracted the spectrum of a
second star which was in the slit and had a magnitude similar to Aql
X-1.  If the velocity errors are due only to statistical effects, one
expects the errors to scale with the  $r$ values like
$ c/(1+r)$ (Tonry \& Davis 1979), with c set such that the
$\chi^2/\nu =1$ under the assumption of a constant velocity for this
star.  If the errors are set by systematic effects, then there should
be no difference in the scatter for high (or low) $r$ values.  In
order to test for systematic effects we first culled out one deviant
velocity ($-700$ \kms) which also had $r<4$.
We then separated the remaining 25 velocities into two
groups based on whether the $r$ values were above or below the median
value.  We find no significant difference in the variance of the
velocity data for the high $r$ value ($r>12$, $\sigma = 28.0$~\kms) or
low $r$ value ($4<r<12$, $\sigma = 29.6$~\kms) groups, and therefore
adapt a uniform error for the Aql X-1 velocities of $\sigma =
29.6$~\kms.

We fit a constant plus a sine wave with a period of 18.9479~hr
(Chevalier \& Ilovaisky 1998) to the 22 Aql X-1 velocities with
correlation coefficients $r>3.0$. The $\chi^2 = 15$, indicating an
acceptable fit, and we find a best fit ${\rm K_c} = 21$~\kms, $\gamma
= 19$~\kms, ${\rm T(0)} = 2450606.8666$~HJD.  Here ${\rm T(0)}$ is
defined as the time for which the sine component of the fit equals
zero, which should occur when the secondary is in front of the
primary.  This best fit sinusoid is plotted in Figure~3.  The 68\%
confidence limits (Lampton, Margon, \& Bowyer 1976) are $5 < \gamma <
32$~\kms, $2 < {\rm K_c} < 41$~\kms, and $2450606.6992 < {\rm T(0)} <
2450606.9755$~HJD, 90\% limits are $ 1 < \gamma < 39$\kms, ${\rm K_c}
< 48$\kms (the phase of T(0) is not bounded within the 90\% limits).
These results are unchanged if we fit instead to the 19.30~hour period
reported by Shahbaz et~al. (1998b).

%
%

We can determine more restrictive limits on ${\rm K_c}$ if we first
limit T(0) by accepting the 18.9479~hr period as orbital and
projecting the time of optical minimum found in section~2.1 forward to
the epoch of our radial velocities, e.g., ${\rm T(0)}=2450606.703
\pm 0.006$~HJD.  Note that this ${\rm T(0)}$ is consistent with that
found above, although one should not assign too much significance to
this since the the radial velocity data alone only provide a $\sim
30$\% restriction on the phase of ${\rm T(0)}$.  These fits show ${\rm
K_c} < 24$~\kms (68\% limits) and ${\rm K_c} < 29$~\kms (90\% limits).

%
%

In order to measure the velocities of the H$\alpha$ line, we fit the
individual spectra to a single Gaussian emission line with a FWHM and
amplitude equal to that measured in the averaged spectrum.  The S/N in
the H$\alpha$ line is high enough that the statistical error in the
fitting procedure is typically $\sim 5$~\kms.  However, as can be seen
in Figure~4, the scatter around the best fit is much larger than
5~\kms, possibly because the intrinsic shape of the line is variable.
In order to estimate errors on the H$\alpha$ velocity curve, we have
re-scaled the errors until $\chi^2/\nu = 1.0$.  Allowing all three
parameters to vary, we then find 68\% error ranges of ${\rm
K}_{H\alpha} = 70 \pm 20$~\kms, $\gamma = 150 \pm 20$~\kms, and ${\rm
T(0)_{H\alpha}} = 2450607.10 \pm 0.05$~HJD.  The 90\% error ranges are
50\% larger.  As before, ${\rm T(0)_{H\alpha}}$ is defined to occur
when the sine component of the the fit equals zero, which in this case
should be when the primary is in front of the secondary, assuming that
the H$\alpha$ velocities accurately trace the motion of the primary
(ie, 0.5~cycles from {\rm T(0)}).

The averaged spectrum of Aql X-1 is shown in Figure~5.  H-Balmer, HeI
and HeII lines are seen in emission.  The H$\alpha$ emission has an
measured FWHM of $845\pm 25$\kms.  Correcting for the instrumental
resolution by subtracting it in quadrature, we estimate intrinsic FWHM
of the line is $830 \pm 25$\kms.  We estimate FWZI~$\sim 2100$~\kms\/ at
H$\alpha$, although this is a lower limit due to the S/N in the
spectrum.  The HeII~4686\AA\/ line does not show the double-peaked
accretion disk profile often seen in LMXB and CVs, and has an
intrinsic FWHM~$=640 \pm 70$~\kms, and a FWZI~$\sim 2000\pm 500$~\kms

We do not find any evidence for a Li~I~6707\AA\/ line in the summed
spectrum.  The noise level in the summed spectrum corresponds to
0.1\AA~EW per resolution element, so we set a 3$\sigma$ upper limit to
the EW of this line of 0.3\AA.  We also note that several photospheric
lines with EW~$\sim 0.2$\AA\/ are readily detectable in the spectrum,
for example the Ca~I~6717\AA\/ line has EW~$ \sim 0.2$\AA.

\section{Discussion:}

{\it What Causes the Observed Modulations?} --- Based on analogy to
other LMXB, the 0.6~mag optical modulation seen during outburst in
1996~July is likely due to reprocessing of x-rays in a
non-axisymmetric disk and/or in the heated face of the companion (van
Paradijs and McClintock 1995).  As both effects create minima at the
same phase, we are not able to distinguish between these two
possibilities.  The relative phasing of the outburst and quiescent
light curves found by Chevalier and Ilovaisky (1998) most likely
supports either of these interpretations.  Additionally, the agreement
between the time of 1996~July optical minimum and the {\rm T(0)} found
from our radial velocity data supports these interpretations, but we
caution the significance of this agreement is low.

Any modulation found during quiescence is likely to be ellipsoidal,
since the K-star is very dominant during quiescence.  However, our
data does not allow us to clearly detect any quiescent modulation.
Shahbaz et~al. (1998b) may detect a quiescent modulation, and are able to
determine a time of minimum light.  Based on the standard model of
x-ray reprocessing during outburst and ellipsoidal modulations during
quiescence, the time of the outburst and quiescent minima should differ by
0.5~cycles.  The accuracy of the period determination of Chevalier and
Ilovaisky (1998) is sufficient to allow meaningful comparison of the
time of minimum light of our outburst modulation to that found in
quiescence (Shahbaz et~al. 1998b).  This difference, $432.212 \pm
0.004$ cycles, is unexpected if the period is orbital in origin.
This implies that either the orbital period or the time of minimum
light in quiescence may be in error.  

{\it What is the Inclination?} --- The empirical relation between the
inclination and the amplitude of outburst modulation found in
$\sim$~dozen LMXB (van Paradijs and McClintock 1995, van Paradijs, van
der Klis, \& Pedersen 1988) leads one to predict an inclination of
$\sim 70^{\rm o}$ from our observed 0.6~mag modulation.  This high
inclination is unexpected given the much lower inclinations implied by
the very small modulations found during quiescence, at I-Band (Shahbaz
et~al. 1998b) and at K-band (this paper).  

	The radial velocity amplitude ${\rm K_c}$ can provide a
measurement of the inclination of the system, if we make the
assumption that the component masses in Aql X-1 are similar to those
in the well studied NS SXT Cen X-4.  This may be a reasonable
assumption, since Cen X-4 has an orbital period and spectral type
(${\rm P_{orb}} = 15.098$~hr, Chevalier et~al. 1989, spectral type
K5-7 V, McClintock and Remillard 1990) similar to Aql X-1.  Previous
studies of Cen X-4 find that ${\rm K_c} = 146$~\kms\/ and $i \sim
40^{\rm o}$ (McClintock and Remillard 1990, Shahbaz, Naylor and
Charles 1993).  Since ${\rm K_c}$ scales directly with sin$(i)$ via
the mass function, the limits we derive from the radial velocity data
alone imply $i = 5^{\rm o} \pm 4^{\rm o}$ (68\%), or $i < 12^{\rm o}$
(90\% confidence limit).  Accepting the 18.9479~hr period as orbital
allows even more restrictive limits of $i < 6^{\rm o}$ (68\%) or $i <
7^{\rm o}$ (90\%).  These limits are much more restrictive than those
placed by the quiescent ellipsoidal modulations, and imply that Aql
X-1 is the lowest inclination SXT known.

{\it What are the Binary Parameters?} --- Optical photometry and
spectroscopy indicate the secondary star in Aql X-1 is of early KV
spectral type (Thorstensen, Charles \& Bowyer 1978, Shahbaz,
et~al. 1997).  Dereddening our IR~magnitudes assuming ${\rm
E(B-V)}=0.4$, we find ${\rm (J-K)_0 = 0.55 \pm 0.21}$, consistent with
G0V--K6V spectral types (Tokunaga 1998).

Secondaries in SXT are generally under massive for their spectral
type; for example, the secondary in Cen X-4 has been estimated to have
a mass ${\rm M_c} \sim 0.1$\msol\/ (McClintock and Remillard 1990), so
for Aql X-1 we expect ${\rm M_c}$\lax~0.8\msol\/ (Allen 1973).
Combining the equation for rotational velocity of the secondary (${\rm
V_{rot}}$) from Wade and Horne (1988) with that for the Roche Lobe
radius from Eggleton (1983) allows one to approximate:

$$ {\rm V_{rot} sin}(i) = 0.52 {\rm K_c} q^{1/3}(1+q)^{2/3} $$

Assuming ${\rm M_x} = 1.4$~\msol, our 90\% upper limit of ${\rm K_c} <
48$~\kms\/ then allows us to approximate ${\rm
V_{rot}sin}(i)$~\lax~27~km~s$^{-1}$.  The more restrictive limits
allowed by using the 18.9479~hr period imply ${\rm
V_{rot}sin}(i)$~\lax~16~km~s$^{-1}$ (90\% confidence).  The one
previously published measurement of the rotational speed of the
secondary finds ${\rm V_{rot}sin}(i) = 62^{+30}_{-20}$~\kms\/ (68\%
limits, Shahbaz et~al. 1997).  This may be consistent with our upper
limits at the $\sim 2\sigma$ level.

The assumptions above imply a mass ratio $q = {\rm M_c /
M_x}$\lax$0.6$.  The mass ratio may also be estimated from the orbital
velocity of the H$\alpha$ line, if we assume that ${\rm K}_{H\alpha} =
{\rm K_x}$.  This yields $q_{H\alpha} = {\rm K}_{H\alpha} / {\rm K_c} =
3.3^{+14.7}_{-2.0}$, or assuming ${\rm K_c} < 48$\kms, $q_{H\alpha} >
0.8$.  While this is inconsistent with the assumed masses for the
components at the 90\% level, it is consistent at the 95\% error level.
One should be wary of mass ratios computed using orbital the velocity
of emission lines, as they may not always correctly indicate the
velocity amplitude of the primary.  Certainly many studies have found
that the phase of the H$\alpha$ radial velocity curve is not that
expected if it accurately tracked the motion of the primary (Orosz
et~al. 1994, Filippenko, Matheson \& Ho 1995).

{\it Caveats} --- As an aside, we note that these results assume that
the period of Aql X-1 is $\sim 19$~hr.  While there is clearly optical
modulation at 19~hrs both in outburst and quiescence, the fact that
the two independent determinations of the period disagree (Chevalier
and Ilovaisky 1998, Shahbaz et~al. 1998b) sheds some doubt on its
orbital origin.

Some of the apparent discrepancies between the inclination as implied in
outburst and in quiescence would be explained if there was 
an `interloper' along the line of sight to Aql X-1, of if the system
was triple. The probability of an interloper
within a  $1''$ radius of Aql X-1 is rather small at 1.6\%, 
based on the density of stars with V$<19$ in the Aql X-1 field.

\subsection{Lithium:}

The non-detection of the Li~6707\AA\/ line in Aql X-1 is interesting
in comparison to the quite high equivalent width ${\rm EW} = 0.480 \pm
0.065$\AA\/ found in Cen X-4 (Martin et~al. 1994, 1996).  Because Li
is quickly destroyed in the stellar interiors of SXT companions, its
detection implies that it is actively being created in these systems.
It may be produced during x-ray outbursts (Martin et~al. 1996) or also
in the hot ADAF during quiescent periods (Yi and Narayan 1997).  In
the latter case, truncation of the ADAF at the Alfven radius decreases
the Li production, and it is necessary to postulate a propeller in
order to eject detectable quantities of Li into the photosphere of the
secondary.  Aql X-1 has recently been found to contain an efficient
propeller consisting of a NS with a ${\rm P_{spin}} = 1.8$~ms and $B \sim
10^8$~G (Zhang et~al. 1998, Campana et~al. 1998).  If Cen X-4 contains
a 31~ms pulsar (Mitsuda 1996) then its Li production should be
much less efficient (Yi and Narayan 1997), and the strength of the
Li~6707\AA\/ line relative to Aql X-1 is difficult to explain. In the
former case, the fact that Aql X-1 has outbursts much more frequently
than all other SXTs would lead one to expect a substantial
Li~6707\AA\/ line, so it's non-detection is once again difficult to
explain.  A possible explanation may lie in the radio observations:
Martin et~al. (1994) postulate that non-thermal radio emission seen in
outbursts of SXT (including Cen X-4, Hjellming and Han 1995) is
evidence of strong particle acceleration, and that these particles
produce Li via spallation.  Despite sensitive radio observations
during numerous Aql X-1 outbursts, radio emission has been detected
only once, at a level of $\sim 1$~mJy during the 1992 outburst
(Hjellming and Han 1995).  More typically radio observations set upper
limits of $\sim 0.5$~mJy (Biggs and Lyne 1996,
Geldzahler 1983).  This may indicate that Aql X-1 is relatively
inefficient at producing non-thermal particles, as scaling the
observed flux of Cen X-4, at $d=1.2$~kpc (McClintock and Remillard
1990), to the distance of Aql X-1 leads one to expect $\sim 2$~mJy
during an outburst.  As a word of caution, we note that Martin
et~al. 1996 show that the strength of the Li~6707\AA\/ line changes
with orbital phase in X-ray Nova Muscae 1996.  Our phase coverage of
Aql X-1 is incomplete, so we cannot discount the possibility that the
line is stronger at phases we were not able to cover.

\acknowledgments
 
We thank Jeff McClintock and an anonymous referee for helpful comments
on earlier drafts of this paper.  This work was supported in part by
the AXAF Science Center through contract NAS8-39073, and NASA grant
NAGW-4269.

\clearpage


\figcaption[aql.vmod.abs.john.ps]{
V-band lightcurve of Aql X-1 from
the FLWO 1.2m telescope during the 1996 July flat-topped x-ray outburst.
A strong modulation at $\sim 19$~hr is clear. Phase 0.0 has 
 been set to the measured time of minimum light ${\rm T(0)} =
2450282.220 \pm 0.003$~HJD.  \label{v.mod.lc}} 

\figcaption[aql.k.abs.kpno.ps]{
 (top): The K$'$-band lightcurve from
the KPNO 2.1m for Aql X-1 and a nearby (presumed constant) star of
similar magnitude. The data have been folded on the 18.9497~hr period
using {\rm T(0)} = 2450282.22~HJD.  (bottom): The lightcurves in
12~phase bins, were the errors are from the measured variations in
each bin.  The ellipsoidal model corresponding to the 90\% upper limit
with q=1 are also plotted. These models have a peak-to-peak amplitude
of $\sim 7\%$ \label{k.lc} }

\figcaption[mike.aql.2bands.fxcor.rvcor.ps]{
The Aql X-1 radial velocities and best fit ($\chi^2/\nu =
15/19$) sinusoid with ${\rm K_c}=21\pm 16$~\kms,  $\gamma = 19\pm
13$~\kms, and {\rm T(0)} = phase 0.0 = 2450606.8666$^{+0.11}_{-0.17}$HJD
(solid line). \label{abs.vel} }

\figcaption[aql.halpha.ps]{
The radial velocities of the Aql X-1 H$\alpha$ emission line
and best fit sinusoid with ${\rm K}_{H\alpha}=70\pm 20$~\kms and $\gamma = 150 \pm
20$~\kms (solid line). 
The scatter around the best fit line is $\sim 55$~\kms;
substantially larger than the typical
measurement error of $\sim 5$~\kms.  \label{halpha.vel}}

\figcaption[aql.ave2.s3.f110.ps]{
The sum total Aql X-1 spectrum from the MMT, taken in June 1997 when
the system was near quiescence.  Total exposure time is 8.7~hours, the
spectral resolution is 3.6\AA.  The EW of H$\alpha = -9.5$\AA, H$\beta =
-4.6$\AA, and HeII~4686\AA$= -6.2$\AA.  
Shown for comparison below the Aql X-1 spectrum
is the spectrum of the star used as a radial velocity template,
GJ~9698, a K8V.  The continuum of this template star has been
artificially modified to approximate that of Aql X-1.\label{opt.spectrum}}


\newpage

%
\centerline{
\psfig{figure=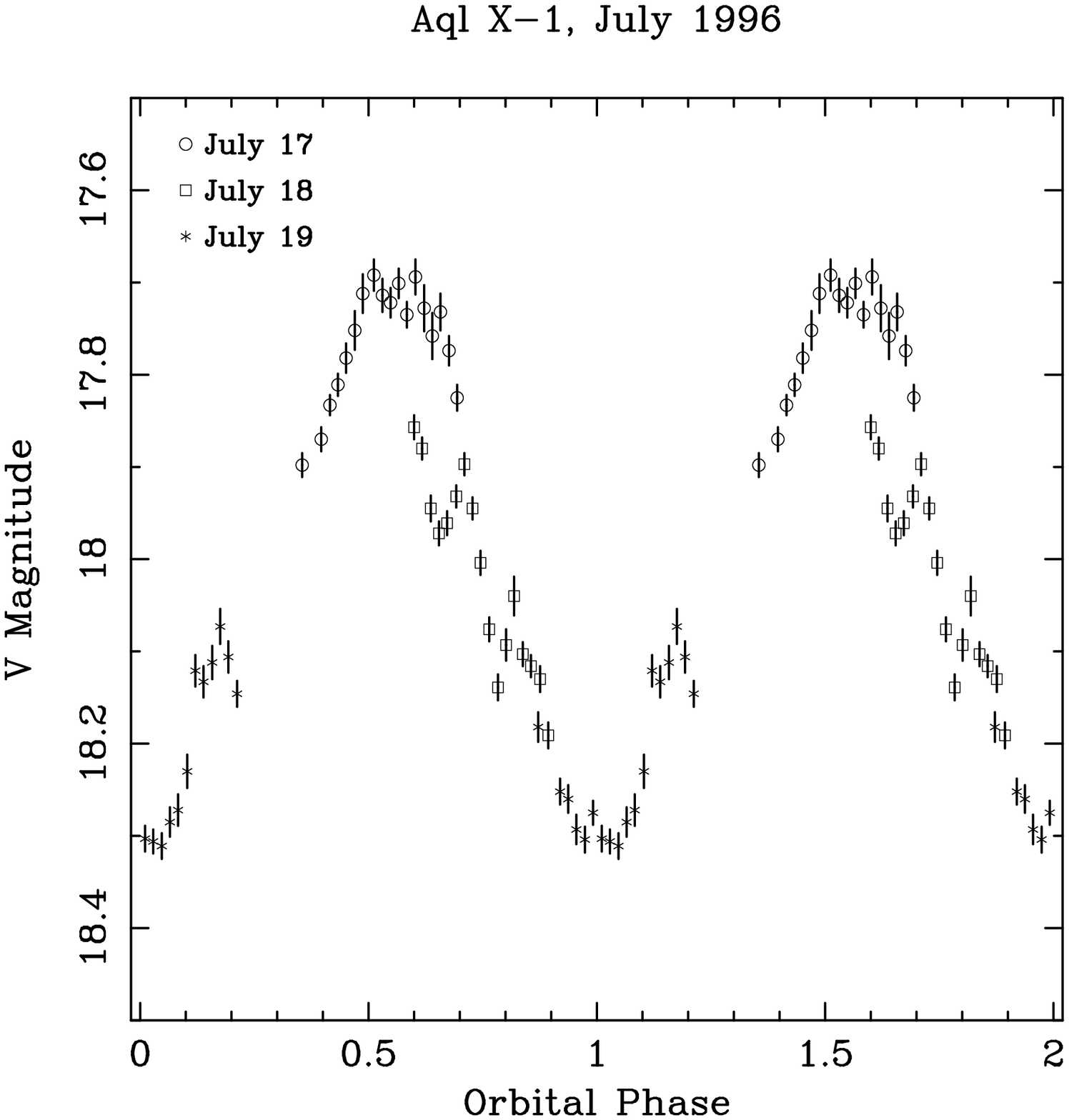,width=6in,angle=-90,width=6in} }
\centerline{
\psfig{figure=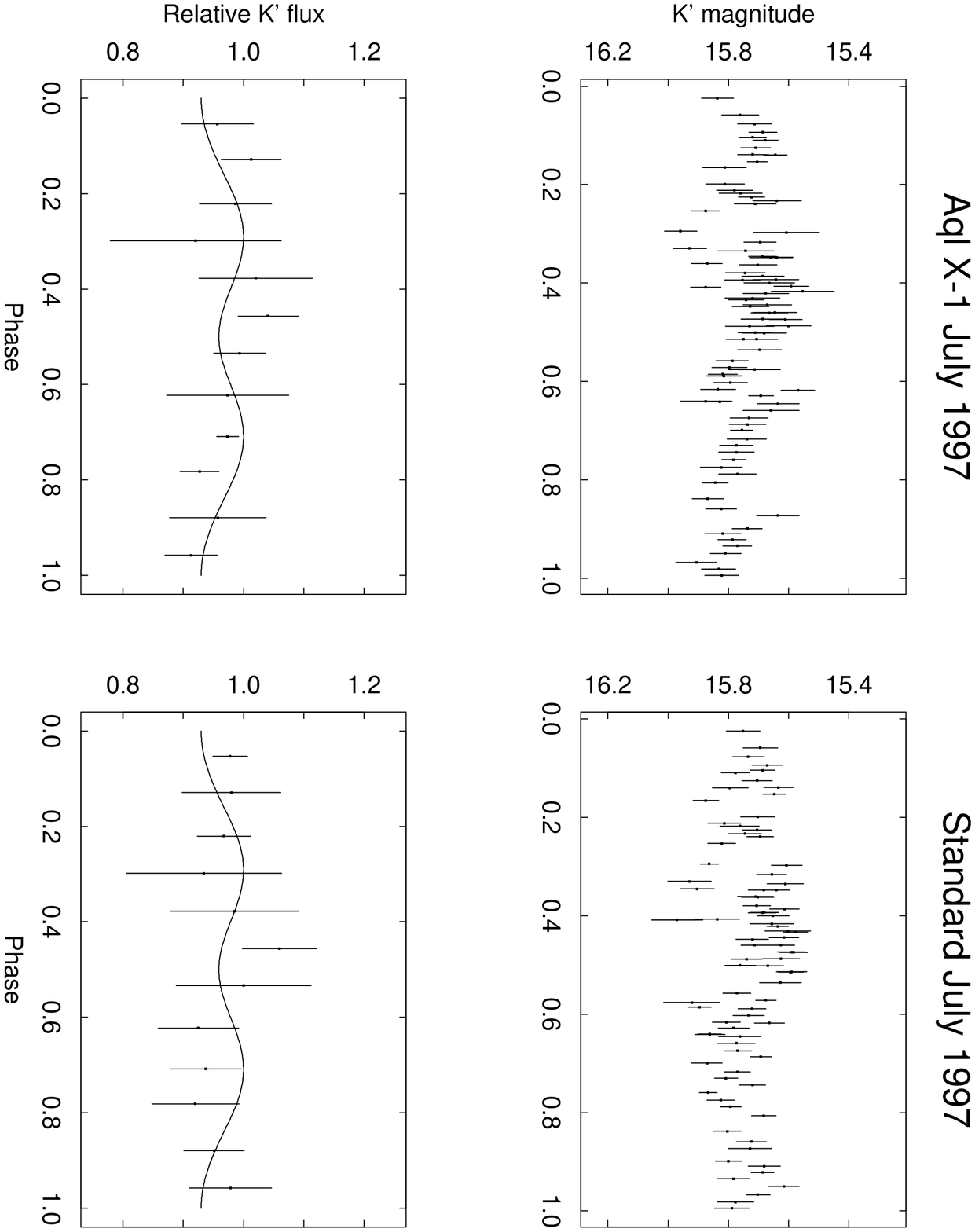,angle=180,width=6in}}
\centerline{
\psfig{figure=apj.fig3.ps,angle=180,width=6in}}
\centerline{
\psfig{figure=apj.fig4.ps,angle=180,width=6in}}
\centerline{
\psfig{figure=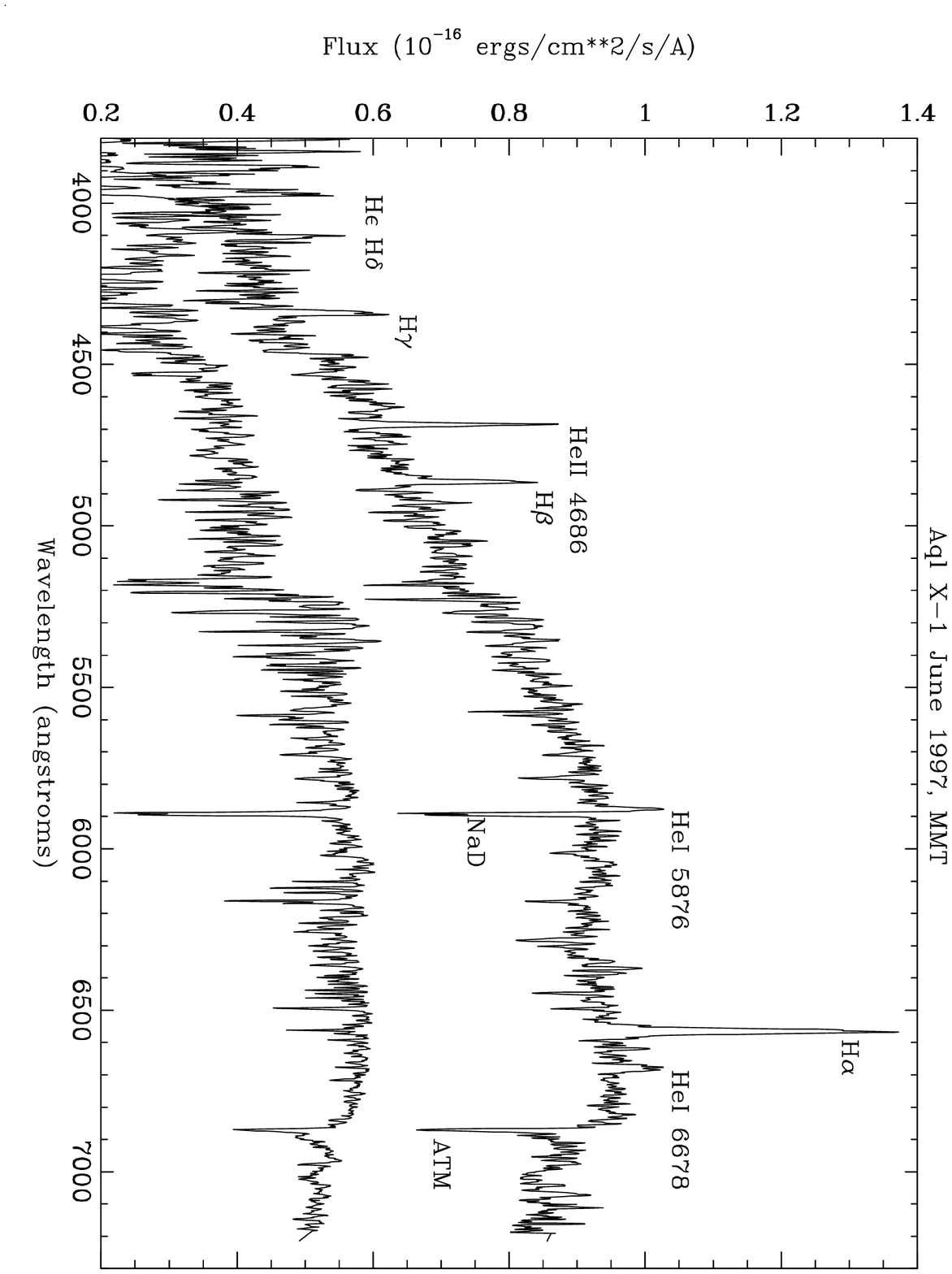}}
%

%

\end{document}